# Density and speed of sound of (iodobenzene + *n*-alkane) liquid mixtures at *T* = (288.15 to 308.15) K. Application of the Prigogine-Flory-Patterson model

Fernando Hevia*[,a], Daniel Lozano-Martín[a], Juan Antonio González[a], Luis Felipe Sanz[a], Isaías García de la Fuente[a], José Carlos Cobos[a]

[a] GETEF. Departamento de Física Aplicada. Facultad de Ciencias. Universidad de Valladolid. Paseo de Belén, 7, 47011 Valladolid, Spain.

* Corresponding author, e-mail: luisfernando.hevia@uva.es

## Abstract

(Iodobenzene + *n*-alkane) liquid mixtures have been studied experimentally, in terms of densities ($\rho$) and speeds of sound ($c$), and theoretically, by the application of the Prigogine-Flory-Patterson (PFP) model. The *n*-alkanes considered are *n*-heptane, *n*-decane, *n*-dodecane, and *n*-tetradecane. $\rho$ and $c$ measurements have been performed at a pressure $p$ = 0.1 MPa and in the temperature range $T$ = (288.15 to 308.15) K. Excess molar volumes ($V_m^E$) and excess isentropic compressibilities ($\kappa_S^E$) have been calculated and correlated by Redlich-Kister polynomials. $\left(\partial V_m^E/\partial T\right)_p$ curves at the same ($p, T$) conditions have been obtained from correlated $V_m^E$ values. The mixtures are characterized by the following features: (i) $V_m^E$ increases with the number of carbon atoms of the *n*-alkane ($n$), being negative for $n$ = 7, small and S-shaped with positive and negative values for $n$ = 10, and positive for $n$ = 12 and $n$ = 14; (ii) $\kappa_S^E$ and $\left(\partial V_m^E/\partial T\right)_p$ are negative, increase with $n$ and are very close to zero for $n$ = 14. From these experimental results and the knowledge of the excess molar enthalpies of mixtures containing fluorobenzene, chlorobenzene or bromobenzene with *n*-alkanes, we have inferred the presence of structural effects, especially important for $n$ = 7 and less relevant for volumetric properties as $n$ increases. The $V_m^E$ obtained are compared with those of other (halogenated benzene + *n*-alkane) liquid mixtures existing in the literature, namely systems with fluorobenzene, chlorobenzene, and bromobenzene. From this comparison and the analysis of the isobaric thermal expansion coefficients of the pure compounds, it is inferred that the interactional effects on $V_m^E$ do not vary appreciably with the length of the *n*-alkane, so the observed $V_m^E$ variation is fundamentally determined by the corresponding variation of the contribution from structural effects. Moreover, the application of the PFP model supports this interpretation, providing free volume contributions to $V_m^E$ that vary parallelly to $V_m^E$ with the length of the *n*-alkane, and interactional contributions that rest approximately constant independently of the *n*-alkane size.

# 1. Introduction

Aromatic halogenated compounds are used in the manufacturing of Li batteries [1], in medicinal chemistry as antidepressant or drugs to reduce cholesterol [2,3], and in the pharmaceutical industry due to the halogen bonding that they can form [4]. They are found in urban air arising from fuel combustion emissions [5]. Thus, it is interesting to study their thermophysical behavior and get a deeper view into their molecular interactions.

In earlier works, we have analyzed mixtures containing a fluorinated benzene and hydrocarbons [6] using thermodynamic properties from the literature and through the application of theoretical models, such as DISQUAC [7], and the concentration-concentration structure factor [8]. In a similar way, we have investigated the thermodynamics of mixtures containing chlorinated or brominated benzenes and alkanes [9]. As the next step, now we turn our attention to systems containing iodobenzene (IB). In this article, we report experimental densities ($\rho$) and speeds of sound ($c$) of (IB + $n$-heptane), (IB + $n$-decane), (IB + $n$-dodecane) and (IB + $n$-tetradecane) liquid mixtures at pressure $p$ = 0.1 MPa and in the temperature range $T$ = (288.15 to 308.15) K. From the measured data, excess molar volumes ($V_m^E$) and excess isentropic compressibilities ($\kappa_S^E$) are calculated and correlated. The temperature derivative of $V_m^E$, $\left(\partial V_m^E / \partial T\right)_p$, is also obtained. Interactions and structural effects in the mixtures are analyzed in terms of their excess properties and by the application of the Prigogine-Flory-Patterson model [10,11].

# 2. Materials and methods

## 2.1. Materials

Pure liquids were used without further purification. Information about their source and purity is displayed in Table 1. Table 2 lists the measured properties of the pure compounds. The comparison of the measured and literature values is fairly good.

## 2.2. Apparatus and procedure

The concentration of the liquid mixtures, given as a mole fraction of IB ($x_1$), was calculated from mass measurements. The masses were determined by weighing using an analytical balance (MSU125P, Sartorius) and correcting for buoyancy effects, with a standard uncertainty of 5·10$^{-5}$ g. Some cautions were taken: (i) to avoid absorption of moisture from air, pure $n$-alkanes were stored with 4 Å molecular sieves after checking that the speed of sound measurements were not affected; (ii) the density of the pure compounds was measured over time to check their stability, staying constant within the uncertainty of the measurements; and (iii) the measurement cells were appropriately filled and closed to avoid partial evaporation.

Pt-100 resistances, calibrated using the triple point of water and the melting point of Ga, were used to measure the temperature of the samples, with a standard uncertainty of 0.01 K.

A densimeter and sound analyzer (DSA 5000, Anton Paar) was employed to determine densities (by the vibrating U-tube method) and speeds of sound (using ultrasonic pulses at 3 MHz center frequency) of the liquid samples with a temperature stability of 0.001 K. More details about the technique and the calibration and test of the apparatus can be found elsewhere [12–14].

## 2.3. Uncertainty

***Uncertainty under repeatability conditions***. The number of digits reported for each of the measured properties has been based on the expanded uncertainty (with a coverage factor of 2) evaluated from the sample standard deviation of measured values obtained under repeatability conditions ($U_1$) [15]. These conditions include the same procedure, the same operators, the same measuring system, the same operating conditions, the same location, and the replicate measurements on samples prepared from the same source liquids over short periods of time. The contribution $U_1$ to the expanded uncertainty is therefore referred to our measurements only. The highest values encountered from measurements performed in this work are $U_1(T)$ = 0.02 K, $U_1(p)$ = 10 kPa, $U_1(x_1)$ = 0.00010, $U_1(c)$ = 0.8 m·s⁻¹, $U_1(\rho)$ = 0.00018 g·cm⁻³ and $U_{1r}(\kappa_S) = [U_{1r}(\rho)^2 + U_{1r}(c)^2]^{1/2}$ for the relative expanded uncertainty ($U_{1r}$) of $\kappa_S$.

***Uncertainty under reproducibility conditions***. For $F = \rho$ and $F = c$ of pure liquids, we have evaluated the root-mean-square deviation $\sigma(F)$ of $N_{\text{lit}}(F)$ literature data points, $F_{\text{lit}}(T)$, from our experimental values, $F_{\text{exp}}(T)$:

$$\sigma(F) = \left[\frac{1}{N_{\text{lit}}(F) - 1} \sum_T \sum_{F_{\text{lit}}(T)} \left(F_{\text{lit}}(T) - F_{\text{exp}}(T)\right)^2\right]^{1/2} \qquad (F = \rho, c) \qquad (1)$$

where the sum is done for all the experimental temperatures $T$ and for every literature data point $F_{\text{lit}}(T)$ at each $T$. These root-mean-square deviations for the pure liquids used in this work have been collected in Table 2. The $\sigma(F)$ values can be considered as estimations of the standard uncertainty evaluated under reproducibility conditions [15], including different locations, operators, measuring systems, measuring procedures, periods of time, and source liquids. Therefore, for pure liquids, the expanded uncertainty (with a coverage factor of 2) evaluated under reproducibility conditions ($U_2$) has been calculated as $U_2(F) = 2\sigma(F)$ and the corresponding relative expanded uncertainty ($U_{2r}$) for $\kappa_S$ as $U_{2r}(\kappa_S) = [U_{2r}(\rho)^2 + U_{2r}(c)^2]^{1/2}$. A conservative estimate of the $U_2(F)$ for the mixtures is chosen to be the maximum value between those of the pure liquids involved (see Table 2).

***Total uncertainty***. The total expanded uncertainty ($U$) should ideally include other factors, such as the purity of the compounds, which is very hard to quantify. For the mole fraction of the mixtures, the total expanded uncertainty has been roughly estimated as $U(x_1)$ = 0.0010. The total uncertainty of $F = \rho$ and $F = c$ can be estimated as (see Table 2):

$$U(F) = [U_1(F)^2 + U_2(F)^2]^{1/2} \qquad (F = \rho, c) \qquad (2)$$

The uncertainties of the excess functions (defined below) are estimated from a statistical analysis of measurements obtained under repeatability conditions and/or comparison with reference values of test systems, such as (cyclohexane + benzene) for the excess molar volume ($V_m^E$) and excess isentropic compressibility ($\kappa_S^E$) [16–18]. The expanded uncertainties ($U$) and relative expanded uncertainties ($U_r$) of $V_m^E$ and $\kappa_S^E$, with a coverage factor of 2, are: $U(V_m^E)$ = 0.010 $|V_m^E|_{\max}$ + 0.005 cm³·mol⁻¹, $U_r(\kappa_S^E)$ = 0.015.

# 3. Equations

If dispersion and absorption of the acoustic wave are negligible, the Newton-Laplace equation allows to determine the isentropic compressibility ($\kappa_S$) from experimental measurements of $\rho$ and $c$:

$$\kappa_S = \frac{1}{\rho c^2} \tag{3}$$

The isothermal compressibility ($\kappa_T$) can be calculated through general thermodynamic relations involving $\kappa_S$, the molar isobaric heat capacity ($C_{pm}$), the molar volume ($V_m$) and the isobaric thermal expansion coefficient ($\alpha_p$):

$$\kappa_T = \kappa_S + \frac{TV_m \alpha_p^2}{C_{pm}} \tag{4}$$

The values $F^{id}$ of the thermodynamic properties of an ideal mixture at the same temperature and pressure as the real mixture are computed from the well-established formulae from Benson and Kiyohara [19–21]:

$$F^{id} = x_1 F_1 + x_2 F_2 \qquad (F = V_m, C_{pm}) \tag{5}$$

$$F^{id} = \phi_1 F_1 + \phi_2 F_2 \qquad (F = \alpha_p, \kappa_T) \tag{6}$$

where $F_i$ is the value of the property $F$ of pure component $i$, $x_i$ represents the mole fraction of component $i$, and $\phi_i = x_i V_{m,i}/V_m^{id}$ is the volume fraction of component $i$. Ideal values of $\kappa_S$ are calculated using the equation [19]:

$$\kappa_S^{id} = \kappa_T^{id} - \frac{TV_m^{id}(\alpha_p^{id})^2}{C_{pm}^{id}} \tag{7}$$

Lastly, the excess quantities $F^E$ are defined by:

$$F^E = F - F^{id} \tag{8}$$

## 4. Results

Some thermodynamic properties ($V_m, \kappa_S, C_{pm}$) of the pure liquids at $p$ = 0.1 MPa are represented as polynomial functions of $T$ of the form:

$$F = \sum_{q=0}^{\ell-1} B_q \left(\frac{T}{K}\right)^q \qquad (F = V_m, \kappa_S, C_{pm}) \tag{9}$$

with $\ell \leq 4$. Parameters for $F = C_{pm}$ are taken from the literature [22–24] and are the only set including $B_3$. For $F = V_m, \kappa_S$, only $(B_0, B_1, B_2)$ are used, and these parameters are obtained by unweighted linear least-squares regressions to experimental values from Table 2. The values of the $B_i$ parameters are listed in Table 3. Smoothed $\alpha_p, \kappa_T$ and $c$ values of the pure compounds can be directly obtained by their definitions and equations (3-4) and (9). The number of parameters used to represent $F = V_m, \kappa_S$ has been chosen so that the calculated values from equation (9) match the experimental values within the uncertainty evaluated under repeatability conditions of measurement.

The experimental $\rho$ and $c$ values of (IB + $n$-alkane) liquid mixtures at $p$ = 0.1 MPa and $T$ = (288.15 to 308.15) K are included in Table 4. The corresponding $\kappa_S$ values can be obtained directly from these using equation (3). The smoothed $\alpha_p$ and $\kappa_T$ values of the pure compounds were used to compute $\kappa_T^{id}$ and $\kappa_S^{id}$ for the mixtures. $V_m^E$ and $\kappa_S^E$ are collected in Table 5 and represented in Figures 1, 2 and S1.

Some excess quantities ($F^E = V_m^E, \kappa_S^E$) are fitted to Redlich-Kister polynomials [25] with $T$-dependent coefficients:

$$F^E = x_1(1-x_1) \sum_{i=0}^{k-1} A_i(T)(2x_1-1)^i \qquad (F^E = V_m^E, \kappa_S^E) \qquad (10)$$

$$A_i(T) = \sum_{j=0}^{m-1} A_{ij} \left(\frac{T}{K}\right)^j \qquad (11)$$

The $A_{ij}$ parameters were obtained by a two-step procedure that ensures the consistency of the obtained values. First, $A_i(T)$ for each property and temperature were obtained by unweighted linear least-squares regressions to equation (10), taking an appropriate number ($k$) of $A_i(T)$ coefficients with guidance from F-tests of additional term [26]. The results are collected in Table S1. Second, $A_i(T)$ values were fitted by unweighted linear least-squares regressions to equation (11), where the number ($m$) of necessary coefficients was chosen so that the $T$-dependence of $F^E$ was correctly represented. For each system, the total number of $A_{ij}$ parameters necessary is between 2 and 4 for $V_m^E$ and between 6 and 8 for $\kappa_S^E$. They are listed in Table 6.

The root-mean-square deviation of the fit to equations (9) and (10-11), $s$, is defined by:

$$s = \left[\frac{1}{N-N_p} \sum_{j=1}^{N} (F_{cal,j} - F_{exp,j})^2\right]^{1/2} \qquad (F = V_m, \kappa_S, C_{pm}, V_m^E, \kappa_S^E) \qquad (12)$$

where $N_p$ stands for the number of $B_i$ parameters of equation (9), or for the total number of $A_i$ parameters in equation (10), or for the total number of $A_{ij}$ parameters of equations (10-11), and $j$ indexes the number $N$ of experimental data points $F_{exp,j}$ and calculated values $F_{cal,j}$. The $s$ values for every fit are included in Tables 3 and 6.

From the mentioned correlations, other properties of the mixtures can be computed as functions of $x_1$ and $T$ using their definitions and equations (3-7), such as $(\partial V_m/\partial T)_p$, $\alpha_p$ or $(\partial V_m^E/\partial T)_p$. In particular, $(\partial V_m^E/\partial T)_p$ are depicted in Figures 3 and S2. For the mixtures, $\kappa_T$ and $\kappa_T^E$ cannot be evaluated because of the lack of $C_{pm}^E$ data.

## 5. Discussion

In this section, if nothing else is specified, the values of the thermophysical properties are considered at $T$ = 298.15 K and $x_1 = 0.5$ (Table 7). As before, we will denote by $n$ the number of carbon atoms of the $n$-alkane.

### 5.1. (IB + *n*-alkane) mixtures

The $V_m^E$ value of (IB + *n*-heptane) is negative, while $V_m^E$ are positive for (IB + *n*-decane), (IB + *n*-dodecane), and (IB + *n*-tetradecane). To the best of our knowledge, there are no data on excess molar enthalpies ($H_m^E$) of the mixtures under study. However, $H_m^E$ are available for other (halogenated benzene + *n*-alkane) systems, such as those containing fluorobenzene [27], chlorobenzene [28,29], and bromobenzene [30], and all of them are positive. Also, for a given *n*-alkane, $H_m^E$(chlorobenzene) < $H_m^E$(bromobenzene) [31], suggesting that $H_m^E$ should be positive for the (IB + *n*-alkane) systems. The negative value of $V_m^E$ for (IB + *n*-heptane) would indicate the presence of important structural effects. This is supported by the negative $(\partial V_m^E/\partial T)_p$ values of (IB + *n*-alkane) systems, larger in absolute value for the *n*-heptane mixture (see below).

Both $V_\mathrm{m}^\mathrm{E}$ and $\left(\partial V_\mathrm{m}^\mathrm{E}/\partial T\right)_p$ of (IB + *n*-alkane) systems increase when *n* increases, suggesting that the positive contribution to $V_\mathrm{m}^\mathrm{E}$ from the breaking of IB-IB interactions by the *n*-alkane gains importance against the negative contribution from structural effects as the length of the *n*-alkane increases.

Negative $\kappa_S^\mathrm{E}$ values are typical of mixtures where structural effects and the formation of interactions between unlike molecules are predominant. In the case of (IB + *n*-alkane) systems, $\kappa_S^\mathrm{E}$ values are negative and increase rapidly as *n* increases. For the *n*-tetradecane mixture, $\kappa_S^\mathrm{E}$ is very close to zero. The same happens with the $\left(\partial V_\mathrm{m}^\mathrm{E}/\partial T\right)_p$ value of this mixture. All these features suggest that the variation of the volumetric properties with *n* is governed by the decrease of the importance of structural effects as *n* increases. This statement is also supported by the application of the Prigogine-Flory-Patterson model (see below).

## 5.2. Comparison with experimental data of other (halogenated benzene + *n*-alkane) mixtures

Now, we shall consider mixtures containing other halogenated benzenes: bromobenzene (BrB), chlorobenzene (ClB) and fluorobenzene (FB). In Figures 4 and 5, respectively, we represent at *T* = 298.15 K and *p* = 0.1 MPa, and as a function of *n*: (i) the difference $(\alpha_{p,1} - \alpha_{p,2})$ between the isobaric thermal expansion coefficient of the pure halogenated benzene ($\alpha_{p,1}$) and the pure *n*-alkane ($\alpha_{p,2}$) (Table S2, [32,33]); and (ii) the $V_\mathrm{m}^\mathrm{E}$ of these systems at equimolar composition (Table S3, [27,34–36]).

On the one hand, the trend observed for $(\alpha_{p,1} - \alpha_{p,2})$ of (IB + *n*-alkane) mixtures, with larger values for shorter *n*-alkanes (Table S2, Figure 4), suggests that the structural effects encountered in (IB + *n*-alkane) mixtures are, at least in part, due to free volume effects. Similar features are observed on (ClB + *n*-alkane) and (BrB + *n*-alkane) mixtures (Figure 4, [31]). On the other hand, Figures 4 and 5 remark the different behavior of (FB + *n*-alkane) systems, where $(\alpha_{p,1} - \alpha_{p,2})$ values are small for shorter *n*-alkanes and grow with *n*, pointing out to stronger structural effects for larger *n*. In contrast, (IB + *n*-alkane), (ClB + *n*-alkane), and (BrB + *n*-alkane) seem to show stronger structural effects for smaller *n*.

In any case, it is interesting to observe that the variation of $(\alpha_{p,1} - \alpha_{p,2})$ with *n* is parallel to the corresponding variation of $V_\mathrm{m}^\mathrm{E}$. This newly suggests that the $V_\mathrm{m}^\mathrm{E}$ variation with *n* is fundamentally determined by the corresponding variation of the relevance of structural effects. Furthermore, for (ClB + *n*-alkane), (BrB + *n*-alkane), and (IB + *n*-alkane) mixtures, the $V_\mathrm{m}^\mathrm{E}$ values are practically the same for *n* = 14. Since, for the latter systems, structural effects have a less relevant weight on $V_\mathrm{m}^\mathrm{E}$, this fact suggests that interactional effects do not change significatively with *n*. These conclusions are also well supported by the application of the Prigogine-Flory-Patterson model, as will be shown.

## 5.3. Prigogine-Flory-Patterson (PFP) model

The Flory model [37–41] is essentially a theory of the van der Waals type which includes free volume and attractive intermolecular interactions. The basic hypotheses of the model are summarized in earlier works [42]. Pure compounds are characterized by two independent reduction parameters, $p^*$ and $V_\mathrm{m}^*$ (reduction pressure and molar core volume, respectively), which can be obtained from $V_\mathrm{m}$, $\alpha_p$, and $\kappa_T$ of the pure compounds. When applied to mixtures, the thermodynamic properties are expressed in terms of one interaction parameter, $X_{12}$, and the most relevant feature is the random mixing hypothesis.

The $X_{12}$ parameter can be easily estimated from one experimental $V_\text{m}^\text{E}$ value using the Flory thermal equation of state and the mixing rules [42]. Patterson and co-workers [10,11] performed two series expansions to gain insight into some excess quantities derived from the model. Thus, to a good approximation, $V_\text{m}^\text{E}$ can be expressed as the sum of three terms:

- The interactional term ($V_\text{m,int}^\text{E}$), proportional to $X_{12}$.
- The curvature term ($V_\text{m,curv}^\text{E}$), always negative and proportional to $-(\bar{V}_1 - \bar{V}_2)^2$, where $\bar{V}_i$ are the reduced volumes of the pure compounds.
- The so-called $p^*$ effect term ($V_{\text{m},p^*\text{eff}}^\text{E}$), which is proportional to $(p_1^* - p_2^*)(\bar{V}_1 - \bar{V}_2)$.

The sum of the curvature and $p^*$ effect terms is called the free volume contribution, $V_\text{m,fv}^\text{E} = V_\text{m,curv}^\text{E} + V_{\text{m},p^*\text{eff}}^\text{E}$. We will not give here the equations resulting from the model, as they are summarized in the cited references.

The PFP model has been applied to gain insight into the volumetric properties of (halogenated benzene + *n*-alkane) systems. Parameters for IB and *n*-alkanes with $n$ = 7,10,12,14 are listed in Table S2; parameters for FB, ClB, BrB, and *n*-alkanes with $n$ = 6,8,9,16 are taken from earlier works [6,9,42]. The interaction parameter $X_{12}$ has been estimated from correlated $V_\text{m}^\text{E}$ values at equimolar composition (for the comparability of the results obtained from the model, the $X_{12}$ values for mixtures containing FB, ClB, and BrB have been estimated in the same way).

The experimental $V_\text{m}^\text{E}$ values and the results from the model are compared in Figure 6. The $X_{12}$ values, together with the different contributions to $V_\text{m}^\text{E}$ at equimolar composition, $T$ = 298.15 K, and $p$ = 0.1 MPa, are collected in Table S3 and represented in Figures 7 and S3.

The interaction parameter $X_{12}$ is proportional to the difference between: (i) the mean intermolecular potential energy of attractive interaction between molecules of the same component, and (ii) the mean intermolecular potential energy of attractive interaction between molecules of different components. The positive $X_{12}$ value of all (halogenated benzene + *n*-alkane) mixtures can be explained by the dipole-dipole interactions among the molecules of the halogenated benzene, whose intensity is higher than the dipole-induced dipole interactions between molecules of the halogenated benzene and the molecules of the *n*-alkane. For $n \geq 7$, its value does not appreciably change with the *n*-alkane size. In fact, it is seen that $V_\text{m,int}^\text{E}$ rests approximately constant as $n$ increases, while $V_\text{m,fv}^\text{E}$ varies parallelly to $V_\text{m}^\text{E}$. We note here the similarity between the $V_\text{m,fv}^\text{E}$ and the $(\alpha_{p,1} - \alpha_{p,2})$ variations with *n*; for mixtures with *n*-alkanes, $V_\text{m,fv}^\text{E}$ is relatively small for (ClB + *n*-alkane), (BrB + *n*-alkane), and (IB + *n*-alkane) mixtures, whereas it has a more significant value for (FB + *n*-alkane). For (IB + *n*-alkane) mixtures, $V_\text{m,fv}^\text{E}$ varies also parallelly to $\left(\partial V_\text{m}^\text{E}/\partial T\right)_p$. All these results are consistent with the analysis carried out above and support our previous statement about the dependence of volumetric properties with *n* being governed by the variation of the importance of structural effects.

Lastly, we compare the experimental results of $\left(\partial V_\text{m}^\text{E}/\partial T\right)_p$ with the ones obtained from the PFP model for (iodobenzene + *n*-alkane) mixtures. We will not consider here other halogenated benzenes for this comparison due to the lack of some experimental data. Values of this property and all the different contributions in the PFP model, $\left(\partial V_\text{m,int}^\text{E}/\partial T\right)_p$, $\left(\partial V_{\text{m},p^*\text{eff}}^\text{E}/\partial T\right)_p$ and $\left(\partial V_\text{m,curv}^\text{E}/\partial T\right)_p$, are listed in Table S4. It is seen that $\left(\partial V_\text{m}^\text{E}/\partial T\right)_p$ is not correctly represented by the model for longer *n*-alkanes. This disagreement is not especially surprising, as the fundamental parameters of this model are temperature-independent and random mixing is assumed. Also, in our estimation of the so-called geometrical parameter of the mixture [42], the molecules of the two components are assumed to have similar shapes and surface areas, a hypothesis that is less correct for higher $n$. However, although there is a quantitative disagreement between the experimental and modeled values of $(\partial V_m^E/\partial T)_p$, we note the qualitative agreement of the variation of the different $(\partial V_m^E/\partial T)_p$ contributions with the above analysis: the interactional contribution remains approximately constant, while the curvature and $p^*$ effect contributions approach zero as $n$ increases.

## 6. Conclusions

Densities and speeds of sound have been measured for (iodobenzene + $n$-alkane) liquid mixtures at $p$ = 0.1 MPa and $T$ = (288.15 to 308.15) K. $V_m^E$ and $\kappa_S^E$ have been calculated and correlated by Redlich-Kister polynomials. $(\partial V_m^E/\partial T)_p$ has been calculated from these correlations. Negative values of $V_m^E$ are found for the $n$-heptane system, which have been attributed to a predominance of structural effects. $V_m^E$ increase with the length of the $n$-alkane chain and become positive for the $n$-decane, $n$-dodecane, and $n$-tetradecane systems; $\kappa_S^E$ and $(\partial V_m^E/\partial T)_p$ are negative and follow the same trend as $V_m^E$ when the $n$-alkane becomes longer. This effect has been attributed to the decrease of the relevance of structural effects as $n$ increases. The comparison with the $V_m^E$ values of mixtures containing $n$-alkanes and other halogenated benzenes (bromobenzene, chlorobenzene, and fluorobenzene) and the application of the PFP model reveal that, in these (halogenated benzene + $n$-alkane) systems, the interactional effects on $V_m^E$ do not vary appreciably with the length of the $n$-alkane, so the observed $V_m^E$ variation is fundamentally determined by the corresponding variation of the contribution from structural effects.

## Funding

This work was supported by Project PID2022-137104NA-I00 funded by MCIN/AEI/10.13039/501100011033/ and by FEDER Una manera de hacer Europa; and Consejería de Educación de Castilla y León, under Project VA100G19 (Apoyo a GIR, BDNS: 425389).

**Table 1**

Sample description.

| Chemical name | CAS Number | Source | [a] Initial purity | Purification method |
|---|---|---|---|---|
| iodobenzene | 591-50-4 | Sigma-Aldrich | 0.999 | none |
| $n$-heptane | 142-82-5 | Fluka | 0.998 | none |
| $n$-decane | 124-18-5 | Sigma-Aldrich | 0.995 | none |
| $n$-dodecane | 112-40-3 | Sigma-Aldrich | 0.998 | none |
| $n$-tetradecane | 629-59-4 | Fluka | 0.995 | none |

[a] In mole fraction. Initial purity certified by the supplier, measured by gas chromatography.

**Table 2**

Thermophysical properties of the pure liquids used in this work at temperature $T$ and pressure $p$ = 0.1 MPa: density ($\rho$), speed of sound ($c$), and isentropic compressibility ($\kappa_S$). $\sigma(F)$, root-mean-square deviation of the literature values from the experimental values, equation (1).

| Compound | [a] Property $F$ | $T$/K | | | | | Literature References | [a] $\sigma(F)$ |
|---|---|---|---|---|---|---|---|---|
| | | 288.15 | 293.15 | 298.15 | 303.15 | 308.15 | | |
| iodobenzene | $\rho$/(g·cm$^{-3}$) | 1.83736 | 1.82972 | 1.82209 | 1.81447 | 1.80685 | [43–47] | 0.002 |
| | $c$/(m·s$^{-1}$) | 1126.8 | 1112.9 | 1099.2 | 1085.9 | 1072.5 | [44,48–50] | 1.1 |
| $n$-heptane | $\rho$/(g·cm$^{-3}$) | 0.68802 | 0.68382 | 0.67960 | 0.67534 | 0.67106 | [b] [51] | 0.00002 |
| | $c$/(m·s$^{-1}$) | 1174.7 | 1152.8 | 1131.2 | 1109.9 | 1088.6 | [50,52–60] | 1.4 |
| $n$-decane | $\rho$/(g·cm$^{-3}$) | 0.73384 | 0.73009 | 0.72631 | 0.72253 | 0.71874 | [b] [61], [62–66] | 0.0007 |
| | $c$/(m·s$^{-1}$) | 1274.7 | 1254.6 | 1234.8 | 1215.2 | 1195.7 | [67–72] | 0.4 |
| $n$-dodecane | $\rho$/(g·cm$^{-3}$) | 0.75257 | 0.74896 | 0.74534 | 0.74172 | 0.73809 | [b] [73], [65,74–76] | 0.0004 |
| | $c$/(m·s$^{-1}$) | 1318.0 | 1298.5 | 1279.2 | 1260.1 | 1241.2 | [58,67,71,77–81] | 0.7 |
| $n$-tetradecane | $\rho$/(g·cm$^{-3}$) | 0.76634 | 0.76281 | 0.75929 | 0.75576 | 0.75223 | [65,82–84] | 0.0006 |
| | $c$/(m·s$^{-1}$) | 1350.3 | 1331.1 | 1312.1 | 1293.4 | 1275.0 | [67,83,85,86] | 0.3 |

[a] Expanded uncertainties ($U_1$), with a coverage factor of 2, evaluated under repeatability conditions of measurement: $U_1(T)$ = 0.02 K, $U_1(p)$ = 10 kPa, $U_1(c)$ = 0.8 m·s$^{-1}$, and $U_1(\rho)$ = 0.00018 g·cm$^{-3}$. Expanded uncertainties ($U_2$), with a coverage factor of 2, evaluated under reproducibility conditions of measurement: for $F = \rho$ and $F = c$, $U_2(F) = 2\sigma(F)$. Total expanded uncertainty ($U$): $U(F) = [U_1(F)^2 + U_2(F)^2]^{1/2}$.

[b] Calculated using REFPROP 10 [87,88].

**Table 3**

Parameters $B_i$ of equation (10) for molar volume ($V_m$), isentropic compressibility ($\kappa_S$), and molar isobaric heat capacity ($C_{pm}$), and root-mean-square deviation of the fit, ($s$, equation (12)), of the pure liquids used in this work in the temperature range $T$ = (288.15 to 308.15) K and pressure $p$ = 0.1 MPa.

| Property | Compound | $B_0$ | $B_1$ | $B_2 \cdot 10^5$ | $B_3 \cdot 10^7$ | $s$ |
|---|---|---|---|---|---|---|
| $V_m$/(cm$^3$·mol$^{-1}$) | iodobenzene | 90.214 | 0.052167 | 6.971 | | 0.0002 |
| | $n$-heptane | 124.02 | -0.026859 | 35.37 | | 0.0008 |
| | $n$-decane | 159.46 | 0.040753 | 27.33 | | 0.0014 |
| | $n$-dodecane | 184.62 | 0.072507 | 25.07 | | 0.0005 |
| | $n$-tetradecane | 209.84 | 0.10234 | 23.55 | | 0.0012 |
| $\kappa_S$/(TPa)$^{-1}$ | iodobenzene | 310.794 | -1.660162 | 718.12 | | 0.08 |
| | $n$-heptane | 3004.32 | -22.64002 | 5507.3 | | 0.22 |
| | $n$-decane | 1493.35 | -10.68131 | 2918.4 | | 0.09 |
| | $n$-dodecane | 1134.12 | -7.832272 | 2273.5 | | 0.02 |
| | $n$-tetradecane | 783.391 | -5.228360 | 1732.9 | | 0.01 |
| [a] $C_{pm}$/(J·mol$^{-1}$·K$^{-1}$) | iodobenzene [23,49,89] | 115.157 | 0.149952 | | | 2.7 |
| | $n$-heptane [22,90,99–106,91–98] | 211.300 | -0.342236 | 151.405 | -7.1543 | 0.30 |
| | $n$-decane [24] | 722.677 | -4.54895 | 1517.30 | -151.24 | 0.92 |
| | $n$-dodecane [24] | 1473.41 | -11.3333 | 3732.70 | -391.15 | 0.86 |
| | $n$-tetradecane [24] | -235.655 | 6.54108 | -2264.44 | 277.94 | 1.3 |

[a] $C_{pm}$ reference parameters recommended from the literature [22–24]. These parameters are valid in the range $T$ = (250 to 320) K for iodobenzene, $T$ = (260 to 400) K for $n$-heptane, $T$ = (247 to 314) K for $n$-decane, $T$ = (267 to 310) K for $n$-dodecane, $T$ = (283 to 310) K for $n$-tetradecane.

**Table 4**

Density ($\rho$) and speed of sound ($c$) of (iodobenzene (1) + $n$-alkane (2)) liquid mixtures as function of the mole fraction of iodobenzene ($x_1$), at temperature $T$ and pressure $p$ = 0.1 MPa. [a]

| $x_1$ | $\rho$/(g·cm$^{-3}$) | | | | | $c$/(m·s$^{-1}$) | | | | |
|---|---|---|---|---|---|---|---|---|---|---|
| | $T$ = 288.15 K | $T$ = 293.15 K | $T$ = 298.15 K | $T$ = 303.15 K | $T$ = 308.15 K | $T$ = 288.15 K | $T$ = 293.15 K | $T$ = 298.15 K | $T$ = 303.15 K | $T$ = 308.15 K |
| (iodobenzene (1) + $n$-heptane (2)) | | | | | | | | | | |
| 0.05629 | 0.73839 | 0.73398 | 0.72954 | 0.72508 | 0.72058 | 1159.2 | 1138.1 | 1117.2 | 1096.6 | 1076.1 |
| 0.09874 | 0.77724 | 0.77267 | 0.76807 | 0.76344 | 0.75879 | 1148.9 | 1128.3 | 1107.9 | 1087.9 | 1067.8 |
| 0.19314 | 0.86688 | 0.86195 | 0.85700 | 0.85203 | 0.84703 | 1129.9 | 1110.3 | 1091.0 | 1071.9 | 1053.0 |
| 0.29294 | 0.96627 | 0.96099 | 0.95567 | 0.95034 | 0.94499 | 1115.0 | 1096.4 | 1078.1 | 1060.1 | 1042.1 |
| 0.39390 | 1.07189 | 1.06624 | 1.06057 | 1.05489 | 1.04918 | 1104.8 | 1087.0 | 1069.6 | 1052.5 | 1035.4 |
| 0.49363 | 1.18191 | 1.17591 | 1.16990 | 1.16387 | 1.15783 | 1099.0 | 1082.1 | 1065.5 | 1049.2 | 1032.9 |
| 0.58767 | 1.29060 | 1.28428 | 1.27795 | 1.27161 | 1.26525 | 1097.4 | 1081.2 | 1065.3 | 1049.6 | 1034.0 |
| 0.69778 | 1.42526 | 1.41857 | 1.41188 | 1.40518 | 1.39848 | 1100.0 | 1084.5 | 1069.2 | 1054.3 | 1039.3 |
| 0.79863 | 1.55516 | 1.54814 | 1.54113 | 1.53411 | 1.52709 | 1106.2 | 1091.2 | 1076.5 | 1062.1 | 1047.7 |
| 0.88830 | 1.67723 | 1.66992 | 1.66263 | 1.65533 | 1.64804 | 1114.2 | 1099.7 | 1085.5 | 1071.6 | 1057.7 |
| 0.94693 | 1.76013 | 1.75265 | 1.74518 | 1.73771 | 1.73024 | 1120.5 | 1106.4 | 1092.4 | 1078.8 | 1065.2 |
| (iodobenzene (1) + $n$-decane (2)) | | | | | | | | | | |
| 0.04653 | 0.76374 | 0.75986 | 0.75596 | 0.75205 | 0.74813 | 1260.7 | 1241.1 | 1221.7 | 1202.5 | 1183.5 |
| 0.09596 | 0.79688 | 0.79286 | 0.78882 | 0.78477 | 0.78072 | 1246.5 | 1227.7 | 1209.0 | 1189.6 | 1170.5 |
| 0.19456 | 0.86761 | 0.86329 | 0.85896 | 0.85462 | 0.85026 | 1221.0 | 1202.5 | 1184.1 | 1165.9 | 1147.9 |
| 0.29276 | 0.94483 | 0.94020 | 0.93556 | 0.93091 | 0.92626 | 1198.2 | 1180.2 | 1162.5 | 1145.1 | 1127.7 |
| 0.39315 | 1.03190 | 1.02693 | 1.02195 | 1.01697 | 1.01198 | 1177.4 | 1160.1 | 1143.0 | 1126.3 | 1109.6 |
| 0.50843 | 1.14375 | 1.13837 | 1.13298 | 1.12758 | 1.12218 | 1157.4 | 1140.8 | 1124.4 | 1108.3 | 1092.3 |
| 0.59557 | 1.23821 | 1.23249 | 1.22675 | 1.22101 | 1.21527 | 1144.8 | 1128.7 | 1112.9 | 1097.4 | 1082.0 |
| 0.69601 | 1.35948 | 1.35334 | 1.34719 | 1.34104 | 1.33489 | 1134.1 | 1118.5 | 1103.1 | 1088.0 | 1073.1 |
| 0.79493 | 1.49435 | 1.48776 | 1.48117 | 1.47458 | 1.46799 | 1126.9 | 1111.9 | 1097.1 | 1082.7 | 1068.3 |
| 0.89813 | 1.65483 | 1.64773 | 1.64064 | 1.63356 | 1.62648 | 1124.8 | 1110.2 | 1096.2 | 1082.3 | 1068.4 |
| 0.94715 | 1.73927 | 1.73192 | 1.72458 | 1.71725 | 1.70993 | 1124.5 | 1110.4 | 1096.5 | 1083.0 | 1069.5 |
| (iodobenzene (1) + $n$-dodecane (2)) | | | | | | | | | | |
| 0.06418 | 0.78768 | 0.78392 | 0.78015 | 0.77637 | 0.77260 | 1299.2 | 1280.4 | 1261.5 | 1242.9 | 1224.4 |
| 0.10602 | 0.81186 | 0.80799 | 0.80412 | 0.80025 | 0.79637 | 1287.9 | 1269.1 | 1250.4 | 1232.0 | 1213.8 |
| 0.19398 | 0.86651 | 0.86242 | 0.85833 | 0.85423 | 0.85012 | 1264.2 | 1245.8 | 1227.7 | 1209.8 | 1192.1 |
| 0.31403 | 0.95046 | 0.94603 | 0.94159 | 0.93714 | 0.93270 | 1233.6 | 1215.9 | 1198.5 | 1181.3 | 1164.3 |
| 0.38937 | 1.00984 | 1.00517 | 1.00049 | 0.99581 | 0.99112 | 1215.4 | 1198.1 | 1181.1 | 1164.4 | 1147.7 |
| 0.49494 | 1.10318 | 1.09814 | 1.09310 | 1.08806 | 1.08300 | 1191.7 | 1175.1 | 1158.7 | 1142.5 | 1126.4 |
| 0.59464 | 1.20479 | 1.19936 | 1.19393 | 1.18850 | 1.18306 | 1171.6 | 1155.5 | 1139.6 | 1124.1 | 1108.6 |
| 0.69262 | 1.32026 | 1.31441 | 1.30855 | 1.30269 | 1.29682 | 1154.3 | 1138.8 | 1123.5 | 1108.5 | 1093.5 |
| 0.79710 | 1.46504 | 1.45866 | 1.45228 | 1.44591 | 1.43953 | 1139.6 | 1124.6 | 1109.9 | 1095.5 | 1081.1 |
| 0.89601 | 1.62852 | 1.62157 | 1.61462 | 1.60769 | 1.60075 | 1130.3 | 1115.9 | 1101.7 | 1087.8 | 1073.9 |
| 0.94745 | 1.72638 | 1.71910 | 1.71182 | 1.70456 | 1.69730 | 1127.7 | 1113.5 | 1099.5 | 1085.9 | 1072.3 |
| (iodobenzene (1) + $n$-tetradecane (2)) | | | | | | | | | | |
| 0.05065 | 0.79009 | 0.78646 | 0.78284 | 0.77921 | 0.77559 | 1336.4 | 1317.5 | 1298.8 | 1280.4 | 1262.1 |
| 0.10535 | 0.81733 | 0.81360 | 0.80986 | 0.80612 | 0.80238 | 1321.8 | 1303.1 | 1284.7 | 1266.5 | 1248.6 |
| 0.19841 | 0.86826 | 0.86431 | 0.86036 | 0.85642 | 0.85247 | 1296.6 | 1278.5 | 1260.5 | 1242.8 | 1225.3 |

| | | | | | | | | | |
|---|---|---|---|---|---|---|---|---|---|
| 0.25033 | 0.89942 | 0.89534 | 0.89127 | 0.88719 | 0.88311 | 1282.3 | 1264.4 | 1246.7 | 1229.3 | 1212.1 |
| 0.39034 | 0.99533 | 0.99087 | 0.98640 | 0.98194 | 0.97747 | 1245.9 | 1228.7 | 1211.8 | 1195.1 | 1178.5 |
| 0.49286 | 1.07924 | 1.07444 | 1.06964 | 1.06484 | 1.06003 | 1219.8 | 1203.1 | 1186.7 | 1170.6 | 1154.6 |
| 0.59492 | 1.17774 | 1.17255 | 1.16736 | 1.16217 | 1.15697 | 1195.0 | 1178.9 | 1163.0 | 1147.5 | 1131.9 |
| 0.69466 | 1.29264 | 1.28700 | 1.28137 | 1.27574 | 1.27009 | 1172.4 | 1156.8 | 1141.5 | 1126.5 | 1111.5 |
| 0.79647 | 1.43485 | 1.42868 | 1.42251 | 1.41634 | 1.41018 | 1152.0 | 1137.0 | 1122.3 | 1107.8 | 1093.4 |
| 0.84951 | 1.52210 | 1.51560 | 1.50910 | 1.50261 | 1.49611 | 1143.0 | 1128.3 | 1113.8 | 1099.6 | 1085.5 |
| 0.89660 | 1.60869 | 1.60186 | 1.59505 | 1.58824 | 1.58143 | 1136.2 | 1121.7 | 1107.5 | 1093.6 | 1079.7 |
| 0.94756 | 1.71369 | 1.70649 | 1.69930 | 1.69212 | 1.68494 | 1130.5 | 1116.3 | 1102.3 | 1088.7 | 1075.1 |

[a] Total expanded uncertainties ($U$) with a coverage factor of 2: $U(T)$ = 0.02 K, $U(p)$ = 10 kPa, $U(x_1)$ = 0.0010. $U(c)$ and $U(\rho)$ are estimated as the maximum $U$ value between those of the pure liquids involved (see Table 2).

**Table 5**

Excess molar volume ($V_m^E$) and excess isentropic compressibility ($\kappa_S^E$) of (iodobenzene (1) + n-alkane (2)) liquid mixtures as function of the mole fraction of iodobenzene ($x_1$), at temperature $T$ and pressure $p$ = 0.1 MPa. [a]

| $x_1$ | $V_m^E$/(cm³·mol⁻¹) | | | | | $\kappa_S^E$/(TPa)⁻¹ | | | | |
|---|---|---|---|---|---|---|---|---|---|---|
| | $T$ = 288.15 K | $T$ = 293.15 K | $T$ = 298.15 K | $T$ = 303.15 K | $T$ = 308.15 K | $T$ = 288.15 K | $T$ = 293.15 K | $T$ = 298.15 K | $T$ = 303.15 K | $T$ = 308.15 K |
| (iodobenzene (1) + n-heptane (2)) | | | | | | | | | | |
| 0.05629 | -0.074 | -0.081 | -0.087 | -0.097 | -0.104 | -19.07 | -20.81 | -22.51 | -24.40 | -26.66 |
| 0.09874 | -0.115 | -0.126 | -0.136 | -0.149 | -0.162 | -31.79 | -34.57 | -37.41 | -40.72 | -44.15 |
| 0.19314 | -0.239 | -0.257 | -0.276 | -0.298 | -0.320 | -55.86 | -60.52 | -65.61 | -70.99 | -77.27 |
| 0.29294 | -0.333 | -0.359 | -0.382 | -0.411 | -0.441 | -74.49 | -80.67 | -87.40 | -94.75 | -102.79 |
| 0.39390 | -0.381 | -0.408 | -0.436 | -0.469 | -0.501 | -86.38 | -93.33 | -101.06 | -109.47 | -118.66 |
| 0.49363 | -0.424 | -0.452 | -0.481 | -0.514 | -0.548 | -91.32 | -98.75 | -106.85 | -115.63 | -125.24 |
| 0.58767 | -0.396 | -0.423 | -0.451 | -0.482 | -0.513 | -89.60 | -96.85 | -104.75 | -113.15 | -122.49 |
| 0.69778 | -0.367 | -0.390 | -0.414 | -0.440 | -0.469 | -79.75 | -86.09 | -92.89 | -100.32 | -108.35 |
| 0.79863 | -0.263 | -0.279 | -0.297 | -0.317 | -0.337 | -62.20 | -66.99 | -72.24 | -77.84 | -84.00 |
| 0.88830 | -0.179 | -0.188 | -0.199 | -0.211 | -0.223 | -39.31 | -42.29 | -45.59 | -49.07 | -52.93 |
| 0.94693 | -0.095 | -0.100 | -0.106 | -0.111 | -0.117 | -20.23 | -21.82 | -23.44 | -25.20 | -27.19 |
| (iodobenzene (1) + n-decane (2)) | | | | | | | | | | |
| 0.04653 | 0.025 | 0.024 | 0.021 | 0.020 | 0.019 | -3.73 | -4.19 | -4.58 | -4.98 | -5.58 |
| 0.09596 | 0.045 | 0.043 | 0.039 | 0.037 | 0.032 | -7.58 | -8.92 | -10.11 | -10.04 | -10.28 |
| 0.19456 | 0.067 | 0.065 | 0.058 | 0.053 | 0.050 | -15.86 | -17.19 | -18.40 | -19.68 | -21.22 |
| 0.29276 | 0.084 | 0.080 | 0.072 | 0.065 | 0.057 | -23.10 | -24.69 | -26.46 | -28.47 | -30.54 |
| 0.39315 | 0.086 | 0.081 | 0.072 | 0.063 | 0.054 | -28.92 | -30.95 | -33.07 | -35.58 | -38.21 |
| 0.50843 | 0.075 | 0.068 | 0.057 | 0.049 | 0.038 | -33.79 | -36.11 | -38.54 | -41.22 | -44.19 |
| 0.59557 | 0.056 | 0.048 | 0.040 | 0.031 | 0.021 | -35.20 | -37.57 | -40.17 | -43.03 | -46.20 |
| 0.69601 | 0.040 | 0.032 | 0.024 | 0.016 | 0.007 | -34.50 | -36.71 | -39.03 | -41.55 | -44.49 |
| 0.79493 | 0.024 | 0.018 | 0.011 | 0.005 | -0.003 | -29.34 | -31.24 | -33.25 | -35.50 | -37.95 |
| 0.89813 | 0.006 | 0.003 | -0.001 | -0.005 | -0.010 | -19.01 | -20.07 | -21.56 | -22.92 | -24.44 |
| 0.94715 | 0.007 | 0.005 | 0.003 | 0.000 | -0.003 | -10.49 | -11.23 | -12.01 | -12.88 | -13.89 |
| (iodobenzene (1) + n-dodecane (2)) | | | | | | | | | | |
| 0.06418 | 0.056 | 0.055 | 0.055 | 0.057 | 0.054 | -1.85 | -2.37 | -2.56 | -2.89 | -3.14 |
| 0.10602 | 0.093 | 0.095 | 0.093 | 0.092 | 0.091 | -3.84 | -4.19 | -4.44 | -4.85 | -5.30 |
| 0.19398 | 0.144 | 0.143 | 0.139 | 0.138 | 0.137 | -7.33 | -7.76 | -8.38 | -9.02 | -9.75 |
| 0.31403 | 0.211 | 0.209 | 0.206 | 0.206 | 0.201 | -11.88 | -12.62 | -13.55 | -14.52 | -15.65 |
| 0.38937 | 0.218 | 0.216 | 0.213 | 0.211 | 0.208 | -14.44 | -15.31 | -16.37 | -17.61 | -18.75 |
| 0.49494 | 0.243 | 0.241 | 0.237 | 0.234 | 0.231 | -17.46 | -18.63 | -19.90 | -21.22 | -22.61 |
| 0.59464 | 0.230 | 0.227 | 0.223 | 0.220 | 0.217 | -19.52 | -20.74 | -22.06 | -23.66 | -25.26 |
| 0.69262 | 0.207 | 0.203 | 0.200 | 0.198 | 0.195 | -19.91 | -21.21 | -22.61 | -24.14 | -25.71 |
| 0.79710 | 0.160 | 0.157 | 0.155 | 0.152 | 0.149 | -17.95 | -19.03 | -20.30 | -21.67 | -23.10 |
| 0.89601 | 0.099 | 0.097 | 0.096 | 0.094 | 0.093 | -12.35 | -13.14 | -13.99 | -14.87 | -15.83 |
| 0.94745 | 0.054 | 0.053 | 0.053 | 0.052 | 0.051 | -7.33 | -7.75 | -8.19 | -8.71 | -9.31 |
| (iodobenzene (1) + n-tetradecane (2)) | | | | | | | | | | |
| 0.05065 | 0.067 | 0.067 | 0.066 | 0.066 | 0.063 | -0.59 | -0.69 | -0.80 | -0.93 | -0.83 |
| 0.10535 | 0.149 | 0.145 | 0.147 | 0.145 | 0.144 | -1.62 | -1.73 | -1.94 | -2.06 | -2.21 |
| 0.19841 | 0.230 | 0.228 | 0.228 | 0.224 | 0.221 | -3.10 | -3.45 | -3.70 | -3.99 | -4.22 |

| | | | | | | | | | |
|---|---|---|---|---|---|---|---|---|---|
| 0.25033 | 0.270 | 0.268 | 0.266 | 0.265 | 0.263 | -3.60 | -3.92 | -4.26 | -4.64 | -4.97 |
| 0.39034 | 0.355 | 0.351 | 0.351 | 0.347 | 0.345 | -6.69 | -7.17 | -7.79 | -8.36 | -8.81 |
| 0.49286 | 0.378 | 0.374 | 0.373 | 0.370 | 0.369 | -8.61 | -9.15 | -9.86 | -10.62 | -11.30 |
| 0.59492 | 0.365 | 0.362 | 0.361 | 0.358 | 0.358 | -10.29 | -10.98 | -11.73 | -12.65 | -13.30 |
| 0.69466 | 0.319 | 0.317 | 0.315 | 0.313 | 0.313 | -11.24 | -11.93 | -12.78 | -13.68 | -14.49 |
| 0.79647 | 0.260 | 0.257 | 0.256 | 0.254 | 0.251 | -10.79 | -11.48 | -12.32 | -13.08 | -13.93 |
| 0.84951 | 0.193 | 0.191 | 0.190 | 0.189 | 0.188 | -9.79 | -10.41 | -11.07 | -11.74 | -12.52 |
| 0.89660 | 0.135 | 0.134 | 0.133 | 0.132 | 0.131 | -8.05 | -8.50 | -9.08 | -9.64 | -10.24 |
| 0.94756 | 0.088 | 0.087 | 0.087 | 0.086 | 0.085 | -4.93 | -5.21 | -5.51 | -5.85 | -6.26 |

[a] Total expanded uncertainties ($U$) and relative expanded uncertainties ($U_r$) with a coverage factor of 2: $U(T)$ = 0.02 K, $U(p)$ = 10 kPa, $U(x_1)$ = 0.0010, $U(V_m^E)$ = 0.010 $|V_m^E|_{max}$ + 0.005 cm$^3 \cdot$mol$^{-1}$, $U_r(\kappa_S^E)$ = 0.015.

**Table 6**

Parameters $A_{ij}$ of equations (10-11) and root-mean-square deviation of the fit (*s*, equation (12)) for excess molar volume ($V_m^E$) and excess isentropic compressibility ($\kappa_S^E$) of (iodobenzene (IB) (1) + *n*-alkane (2)) liquid mixtures used in this work in the temperature range $T$ = (288.15 to 308.15) K and pressure $p$ = 0.1 MPa. $N$ stands for the number of experimental points used for the fit.

| Property | System | $A_{00}$ | $A_{01}\cdot 10^2$ | $A_{02}\cdot 10^4$ | $A_{10}$ | $A_{11}\cdot 10^2$ | $A_{12}\cdot 10^2$ | $A_{20}$ | $A_{21}$ | $A_{30}$ | $N$ | $s$ |
|---|---|---|---|---|---|---|---|---|---|---|---|---|
| $V_m^E$ /(cm³·mol⁻¹) | IB + *n*-heptane | -13.047 | 9.953 | -2.0808 | | | | | | | 55 | 0.015 |
| | IB + *n*-decane | 2.3736 | -0.7209 | | 0.1913 | -0.156 | | | | | 55 | 0.002 |
| | IB + *n*-dodecane | 1.6802 | -0.2491 | | | | | | | | 55 | 0.005 |
| | IB + *n*-tetradecane | 2.0562 | -0.1922 | | | | | | | | 60 | 0.008 |
| $\kappa_S^E$/(TPa)⁻¹ | IB + *n*-heptane | -3548.4 | 2769.9 | -577.95 | -23.500 | | | 34.1 | -0.185 | | 55 | 0.09 |
| | IB + *n*-decane | -1585.4 | 1168.0 | -230.55 | -1974.2 | 1355 | -2.4056 | 79.1 | -0.355 | | 55 | 0.33 |
| | IB + *n*-dodecane | -581.45 | 440.77 | -91.424 | 148.92 | -68.25 | | 110 | -0.465 | -17 | 55 | 0.08 |
| | IB + *n*-tetradecane | 122.45 | -54.434 | | 97.486 | -46.37 | | 68.0 | -0.323 | -14 | 60 | 0.17 |

**Table 7**

Excess molar volume ($V_m^E$), temperature derivative of excess molar volume $\left(\partial V_m^E/\partial T\right)_p$, and excess isentropic compressibility ($\kappa_S^E$) of (iodobenzene (1) + $n$-alkane (2)) liquid mixtures at equimolar composition, at temperature $T$ = 298.15 K and pressure $p$ = 0.1 MPa.

| System | $V_m^E$/(cm³·mol⁻¹) | $\left(\partial V_m^E/\partial T\right)_p$/(10⁻³·cm³·mol⁻¹·K⁻¹) | $\kappa_S^E$/(TPa)⁻¹ |
|---|---|---|---|
| iodobenzene + $n$-heptane | -0.467 | -6.14 | -106.88 |
| iodobenzene + $n$-decane | 0.056 | -1.80 | -38.11 |
| iodobenzene + $n$-dodecane | 0.234 | -0.62 | -20.00 |
| iodobenzene + $n$-tetradecane | 0.371 | -0.48 | -9.95 |

**Figure 1**

Excess molar volume ($V_m^E$) of (iodobenzene (1) + $n$-alkane (2)) liquid mixtures as a function of the iodobenzene mole fraction ($x_1$) at temperature $T$ = 298.15 K and pressure $p$ = 0.1 MPa. Symbols, experimental values: (□) $n$-heptane, (◇) $n$-decane, (△) $n$-dodecane, (×) $n$-tetradecane. Solid lines are calculated with equations (10-11) using the coefficients from Table 6.

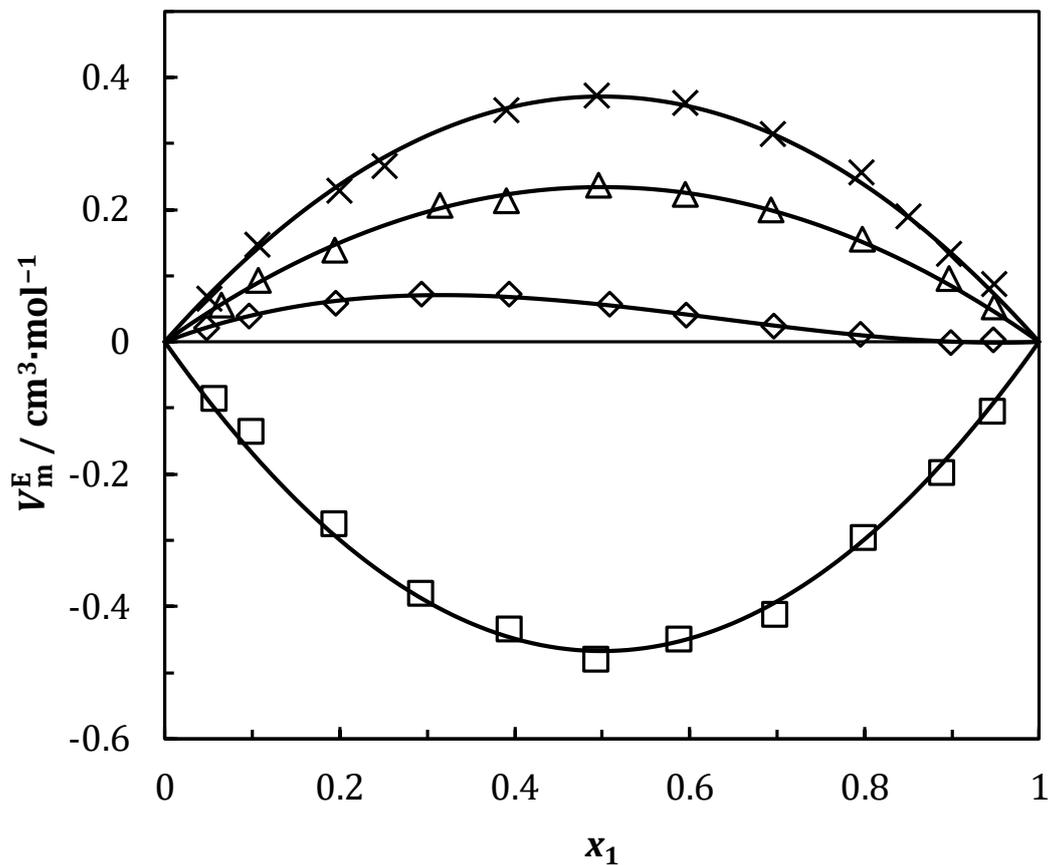

**Figure 2**

Excess isentropic compressibility ($\kappa_S^E$) of (iodobenzene (1) + *n*-alkane (2)) liquid mixtures as a function of the iodobenzene mole fraction ($x_1$) at temperature $T$ = 298.15 K and pressure $p$ = 0.1 MPa. Symbols, experimental values: (□) *n*-heptane, (◇) *n*-decane, (△) *n*-dodecane, (×) *n*-tetradecane. Solid lines are calculated with equations (10-11) using the coefficients from Table 6.

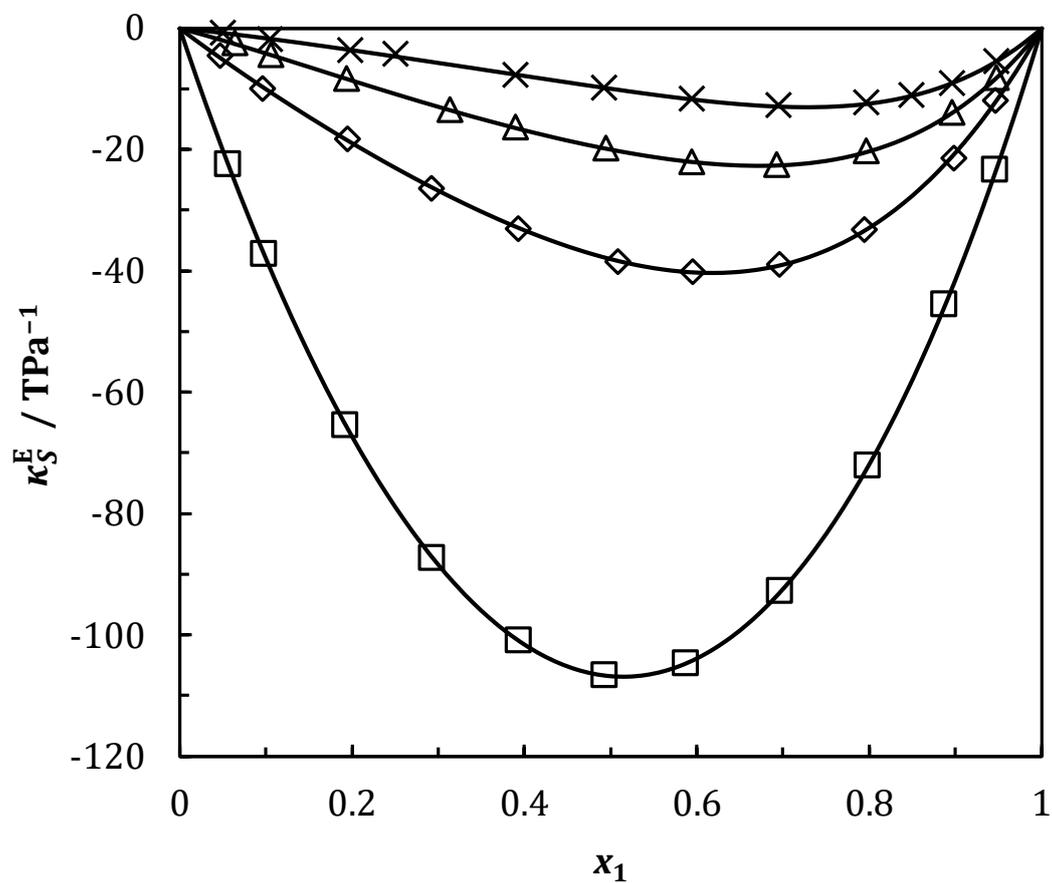

**Figure 3**

Smoothed curves of the temperature derivative of excess molar volumes $(\partial V_m^E/\partial T)_p$ of (iodobenzene (1) + n-alkane (2)) liquid mixtures as a function of the iodobenzene mole fraction ($x_1$) at temperature $T$ = 298.15 K and pressure $p$ = 0.1 MPa: (solid line) *n*-heptane, (dotted line) *n*-decane, (short-dashed line) *n*-dodecane, (long-dashed line) *n*-tetradecane. The calculations were performed from equations (10-11) using the coefficients from Table 6.

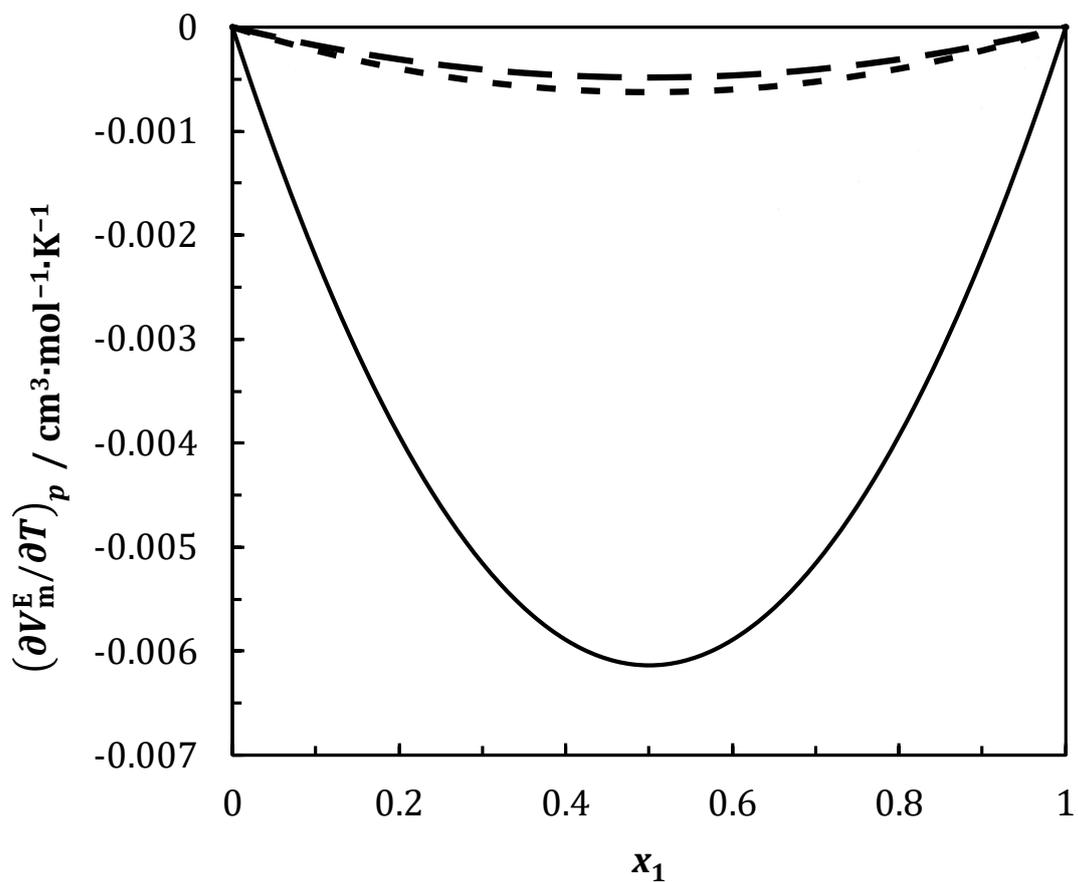

**Figure 4**

Isobaric thermal expansion coefficient difference $(\alpha_{p,1} - \alpha_{p,2})$ between the pure halogenated benzene (1) and the pure *n*-alkane (2) as a function of the number of carbon atoms of the *n*-alkane (*n*) at temperature $T$ = 298.15 K and pressure $p$ = 0.1 MPa. Symbols: (×) fluorobenzene, (△) chlorobenzene, (◇) bromobenzene, (□) iodobenzene. For references see Table S2 and [32,33]. Dashed lines are drawn to aid the eye of the reader.

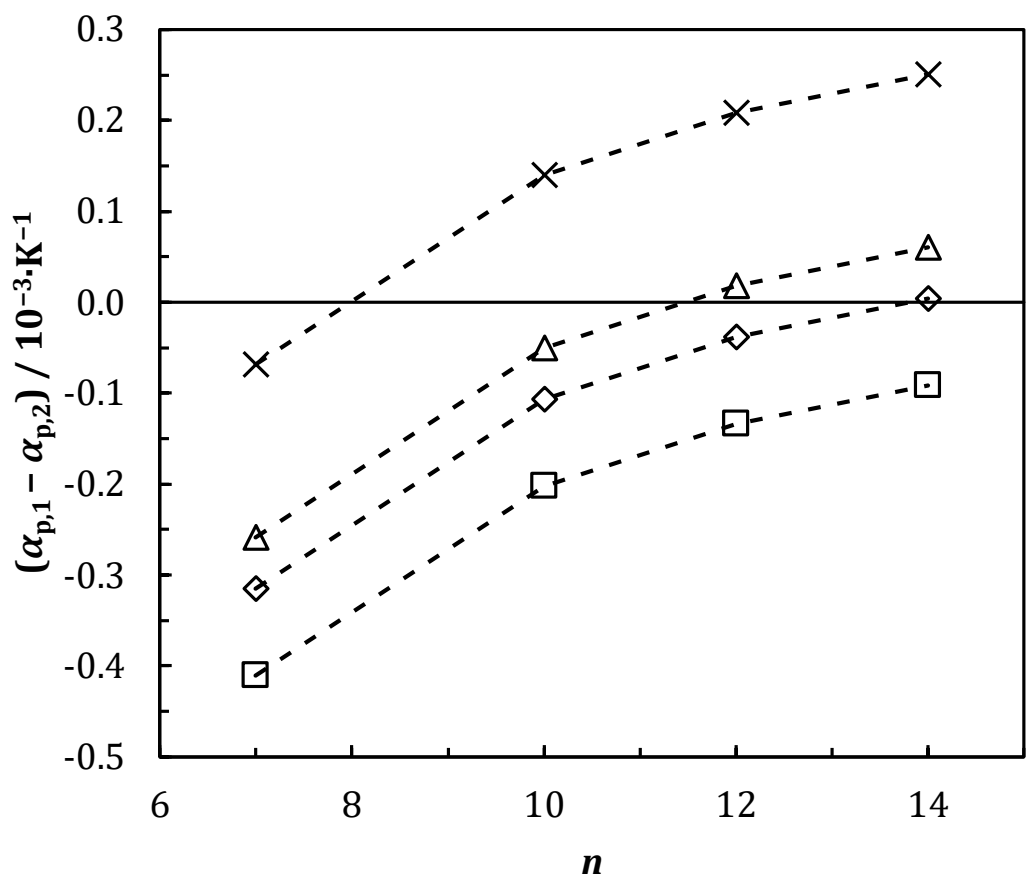

**Figure 5**

Excess molar volume ($V_m^E$) of (halogenated benzene (1) + $n$-alkane (2)) liquid mixtures as a function of the number of carbon atoms of the $n$-alkane ($n$), at equimolar composition, temperature $T$ = 298.15 K and pressure $p$ = 0.1 MPa. Symbols: (×) fluorobenzene, (△) chlorobenzene, (◇) bromobenzene, (□) iodobenzene. For references see Table S3. Dashed lines are drawn to aid the eye of the reader.

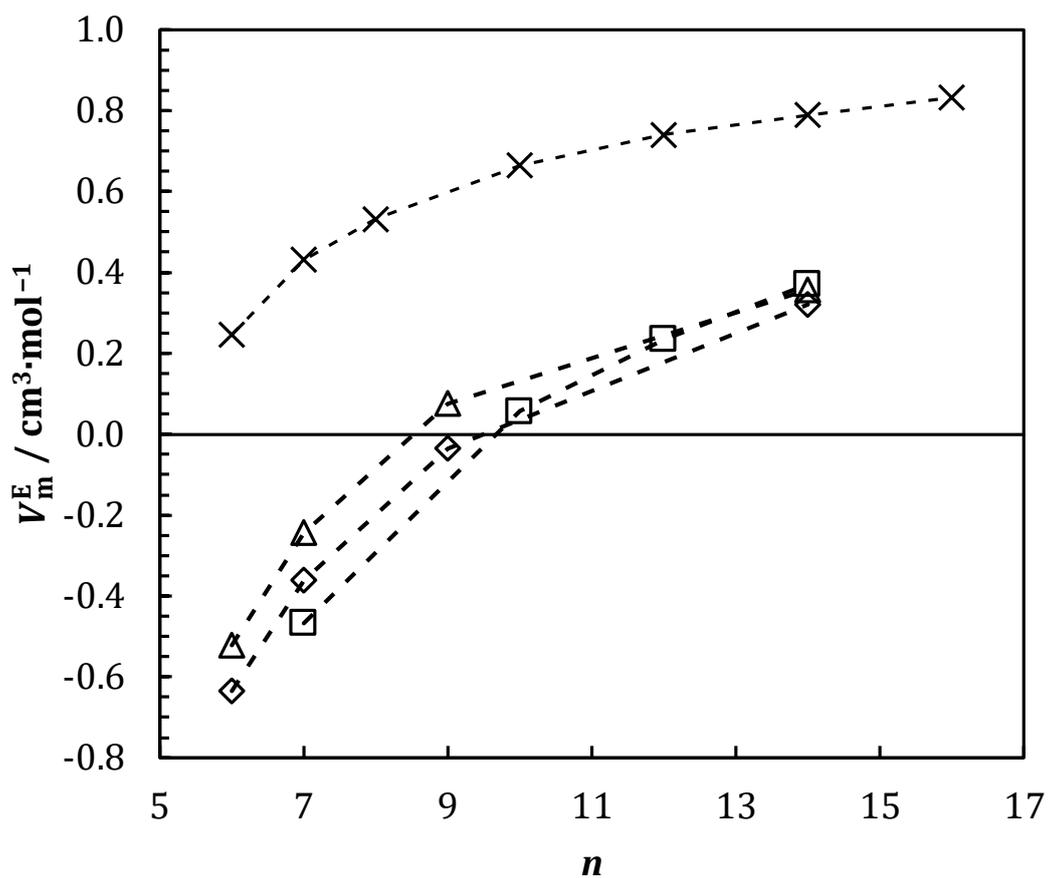

**Figure 6**

Prigogine-Flory-Patterson (PFP) model results on the excess molar volume ($V_m^E$) of (iodobenzene (1) + $n$-alkane (2)) liquid mixtures as a function of the iodobenzene mole fraction ($x_1$) at temperature $T$ = 298.15 K and pressure $p$ = 0.1 MPa. Symbols, experimental values: (□) $n$-heptane, (◇) $n$-decane, (△) $n$-dodecane, (×) $n$-tetradecane. Solid lines, PFP calculations.

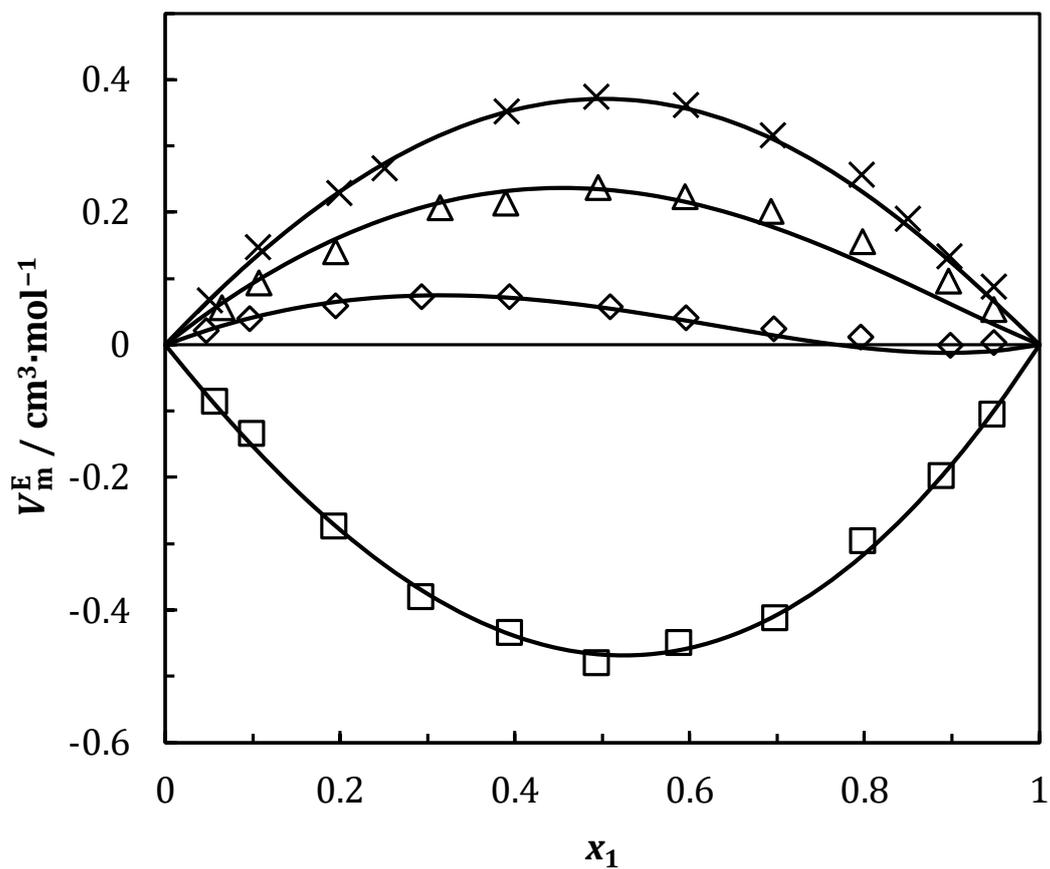

**Figure 7**

Contributions to the excess molar volume ($V_m^E$), obtained from the application of the Prigogine-Flory-Patterson (PFP) model to (halogenated benzene (1) + $n$-alkane (2)) liquid mixtures at equimolar composition, temperature $T$ = 298.15 K and pressure $p$ = 0.1 MPa, as functions of the number of carbon atoms of the $n$-alkane ($n$). Dashed lines, interactional term ($V_{m,int}^E$); solid lines, free volume term ($V_{m,fv}^E = V_{m,curv}^E + V_{m,p^*eff}^E$, sum of the curvature term and the $p^*$ effect term). Symbols: (×) fluorobenzene, (△) chlorobenzene, (◇) bromobenzene, (□) iodobenzene. For references, see Table S3.

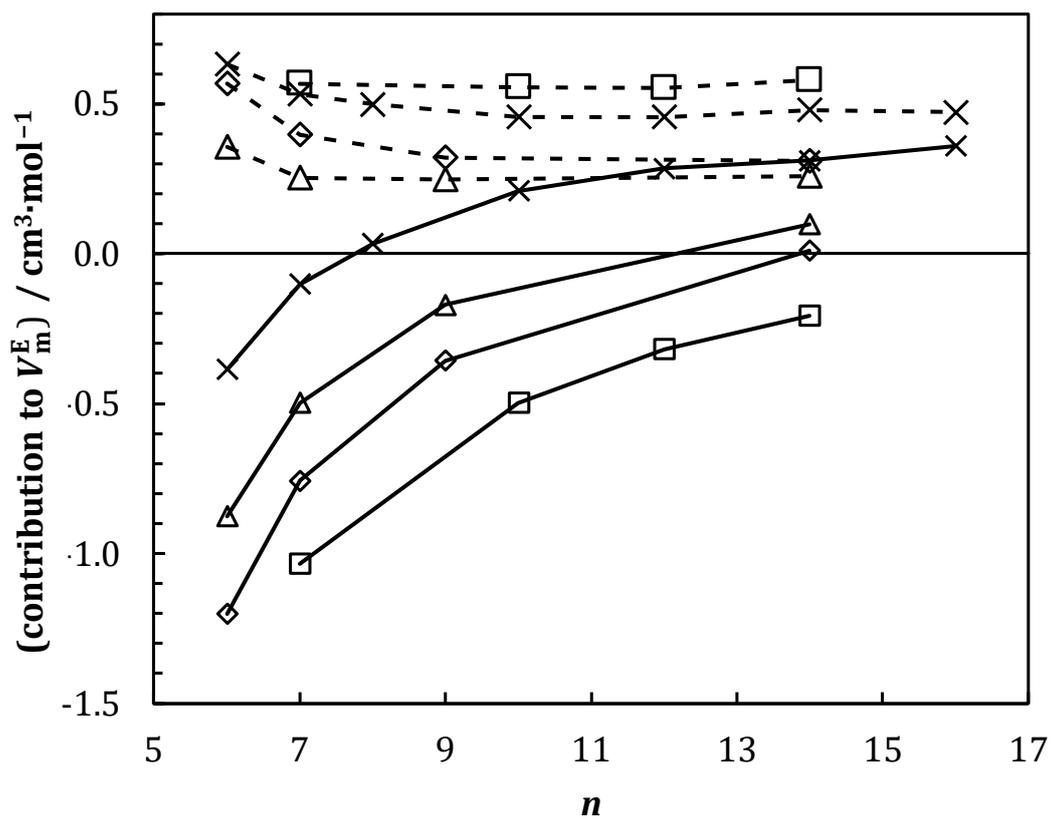

# Density and speed of sound of (iodobenzene + *n*-alkane) liquid mixtures at *T* = (288.15 to 308.15) K. Application of the Prigogine-Flory-Patterson model

# Supplementary Material

Fernando Hevia*,a, Daniel Lozano-Martína, Juan Antonio Gonzáleza, Luis Felipe Sanza, Isaías García de la Fuentea, José Carlos Cobosa

a G.E.T.E.F. Departamento de Física Aplicada. Facultad de Ciencias. Universidad de Valladolid. Paseo de Belén, 7, 47011 Valladolid, Spain.

* Corresponding author, e-mail: luisfernando.hevia@uva.es

**Table S1**

Parameters $A_i$ of equation (10) and root-mean-square deviation of the fit, (s, equation (12)) for excess molar volume ($V_m^E$) and excess isentropic compressibility ($\kappa_S^E$) of (iodobenzene (IB) (1) + n-alkane (2)) liquid mixtures used in this work in the temperature range $T$ = (288.15 to 308.15) K and pressure $p$ = 0.1 MPa.

| Property | System | T/K | $A_0$ | $A_1$ | $A_2$ | $A_3$ | s |
|---|---|---|---|---|---|---|---|
| $V_m^E$ /cm³·mol⁻¹ | IB + n-heptane | 288.15 | -1.64 | | | | 0.017 |
| | | 293.15 | -1.75 | | | | 0.015 |
| | | 298.15 | -1.87 | | | | 0.015 |
| | | 303.15 | -2.00 | | | | 0.014 |
| | | 308.15 | -2.13 | | | | 0.014 |
| | IB + n-decane | 288.15 | 0.292 | -0.26 | | | 0.003 |
| | | 293.15 | 0.265 | -0.27 | | | 0.003 |
| | | 298.15 | 0.225 | -0.27 | | | 0.003 |
| | | 303.15 | 0.190 | -0.28 | | | 0.003 |
| | | 308.15 | 0.150 | -0.29 | | | 0.002 |
| | IB + n-dodecane | 288.15 | 0.962 | | | | 0.006 |
| | | 293.15 | 0.951 | | | | 0.006 |
| | | 298.15 | 0.937 | | | | 0.006 |
| | | 303.15 | 0.926 | | | | 0.006 |
| | | 308.15 | 0.912 | | | | 0.005 |
| | IB + n-tetradecane | 288.15 | 1.50 | | | | 0.008 |
| | | 293.15 | 1.49 | | | | 0.008 |
| | | 298.15 | 1.49 | | | | 0.008 |
| | | 303.15 | 1.47 | | | | 0.008 |
| | | 308.15 | 1.46 | | | | 0.008 |
| $\kappa_S^E$ /(TPa)⁻¹ | IB + n-heptane | 288.15 | -365.7 | -23.7 | -19.1 | | 0.09 |
| | | 293.15 | -395.2 | -23.8 | -20.8 | | 0.08 |
| | | 298.15 | -427.7 | -23.9 | -21.0 | | 0.08 |
| | | 303.15 | -462.7 | -23.5 | -21.6 | | 0.10 |
| | | 308.15 | -501.1 | -22.6 | -23.4 | | 0.11 |
| | IB + n-decane | 288.15 | -134.0 | -67.3 | -20.9 | | 0.4 |
| | | 293.15 | -142.8 | -69.3 | -22.6 | | 0.2 |
| | | 298.15 | -152.2 | -72.6 | -30.3 | | 0.3 |
| | | 303.15 | -163.4 | -77.6 | -28.3 | | 0.3 |
| | | 308.15 | -175.4 | -83.0 | -28.5 | | 0.4 |
| | IB + n-dodecane | 288.15 | -70.53 | -47.5 | -24.3 | -19 | 0.13 |
| | | 293.15 | -74.89 | -51.6 | -27.0 | -16 | 0.08 |
| | | 298.15 | -79.97 | -54.3 | -28.8 | -17 | 0.05 |
| | | 303.15 | -85.58 | -58.1 | -31.0 | -16 | 0.05 |
| | | 308.15 | -91.28 | -61.3 | -34.0 | -17 | 0.08 |
| | IB + n-tetradecane | 288.15 | -34.58 | -36.3 | -25.1 | -13 | 0.18 |
| | | 293.15 | -36.90 | -38.2 | -26.9 | -13 | 0.18 |
| | | 298.15 | -39.71 | -40.6 | -28.6 | -14 | 0.19 |
| | | 303.15 | -42.78 | -43.4 | -39.6 | -13 | 0.19 |
| | | 308.15 | -45.25 | -45.3 | -31.8 | -16 | 0.19 |

**Table S2**

Pure compound parameters used for the application of the Prigogine-Flory-Patterson model to (iodobenzene (1) + $n$-alkane (2)) liquid mixtures used in this work at temperature $T$ = 298.15 K and pressure $p$ = 0.1 MPa: molar volume ($V_\text{m}$), isobaric thermal expansion coefficient ($\alpha_p$), isothermal compressibility ($\kappa_T$), molar core volume ($V_\text{m}^*$), and reduction pressure ($p^*$).

| Compound | $V_\text{m}$/(cm³·mol⁻¹) | $\alpha_p$/(10⁻³K⁻¹) | $\kappa_T$/(TPa)⁻¹ | $V_\text{m}^*$/(cm³·mol⁻¹) | $p^*$/MPa |
|---|---|---|---|---|---|
| iodobenzene | 111.96 | 0.837 | 600.58 | 92.28 | 611.5 |
| $n$-heptane | 147.44 | 1.248 | 1454.49 | 113.72 | 429.9 |
| $n$-decane | 195.90 | 1.040 | 1103.92 | 155.98 | 442.9 |
| $n$-dodecane | 228.53 | 0.971 | 991.01 | 184.04 | 450.3 |
| $n$-tetradecane | 261.28 | 0.929 | 918.44 | 211.91 | 458.3 |

**Table S3**

Contributions to the excess molar volume ($V_m^E$), obtained from the application of the Prigogine-Flory-Patterson model to (halogenated benzene (1) + $n$-alkane (2)) liquid mixtures at equimolar composition, temperature $T$ = 298.15 K and pressure $p$ = 0.1 MPa. The halogenated benzenes are: fluorobenzene (FB), chlorobenzene (ClB), bromobenzene (BrB) and iodobenzene (IB). $X_{12}$, interaction parameter estimated from correlated $V_m^E$ values at the mentioned conditions; $V_{m,\text{int}}^E$, interactional term; $V_{m,p^*\text{eff}}^E$, $p^*$ effect term; $V_{m,\text{curv}}^E$, curvature term.

| System | $V_m^E$ /(cm$^3$·mol$^{-1}$) | $X_{12}$ /MPa | $V_{m,\text{int}}^E$ /(cm$^3$·mol$^{-1}$) | $V_{m,p^*\text{eff}}^E$ /(cm$^3$·mol$^{-1}$) | $V_{m,\text{curv}}^E$ /(cm$^3$·mol$^{-1}$) |
|---|---|---|---|---|---|
| FB + $n$-hexane | 0.246 [1] | 29.93 | 0.632 | -0.337 | -0.049 |
| FB + $n$-heptane | 0.431 [1] | 26.53 | 0.532 | -0.095 | -0.006 |
| FB + $n$-octane | 0.532 [1] | 25.27 | 0.498 | 0.034 | 0.000 |
| FB + $n$-decane | 0.664 [1] | 23.73 | 0.455 | 0.241 | -0.032 |
| FB + $n$-dodecane | 0.740 [1] | 24.24 | 0.456 | 0.363 | -0.079 |
| FB + $n$-tetradecane | 0.789 [1] | 25.83 | 0.479 | 0.431 | -0.121 |
| FB + $n$-hexadecane | 0.832 [1] | 25.83 | 0.472 | 0.540 | -0.180 |
| ClB + $n$-hexane | -0.521 [2] | 17.27 | 0.356 | -0.675 | -0.202 |
| ClB + $n$-heptane | -0.243 [2] | 12.89 | 0.254 | -0.400 | -0.097 |
| ClB + $n$-nonane | 0.076 [2] | 12.82 | 0.247 | -0.156 | -0.015 |
| ClB + $n$-tetradecane | 0.356 [3] | 13.74 | 0.259 | 0.105 | -0.008 |
| BrB + $n$-hexane | -0.634 [4] | 28.89 | 0.568 | -0.927 | -0.275 |
| BrB + $n$-heptane | -0.361 [4] | 20.90 | 0.397 | -0.607 | -0.151 |
| BrB + $n$-nonane | -0.036 [4] | 17.17 | 0.321 | -0.315 | -0.042 |
| BrB + $n$-tetradecane | 0.320 [4] | 16.61 | 0.309 | 0.011 | 0.000 |
| IB + $n$-heptane | -0.467 | 29.64 | 0.567 | -0.761 | -0.273 |
| IB + $n$-decane | 0.056 | 29.13 | 0.555 | -0.413 | -0.086 |
| IB + $n$-dodecane | 0.234 | 28.97 | 0.553 | -0.278 | -0.041 |
| IB + $n$-tetradecane | 0.371 | 30.29 | 0.579 | -0.187 | -0.021 |

**Table S4**

Contributions to the temperature derivative of the excess molar volume ($(\partial V_m^E/\partial T)_p$), obtained experimentally and from the application of the Prigogine-Flory-Patterson (PFP) model to (iodobenzene (IB) (1) + $n$-alkane (2)) liquid mixtures at equimolar composition, temperature $T$ = 298.15 K and pressure $p$ = 0.1 MPa. $(\partial V_{m,\text{int}}^E/\partial T)_p$, interactional term; $(\partial V_{m,p^*\text{eff}}^E/\partial T)_p$ effect term; $(\partial V_{m,\text{curv}}^E/\partial T)_p$ curvature term.

| System | $(\partial V_m^E/\partial T)_p$ /($10^{-3}\cdot$cm$^3\cdot$mol$^{-1}\cdot$K$^{-1}$) | | $(\partial V_{m,\text{int}}^E/\partial T)_p$ /($10^{-3}\cdot$cm$^3\cdot$mol$^{-1}\cdot$K$^{-1}$) | $(\partial V_{m,p^*\text{eff}}^E/\partial T)_p$ /($10^{-3}\cdot$cm$^3\cdot$mol$^{-1}\cdot$K$^{-1}$) | $(\partial V_{m,\text{curv}}^E/\partial T)_p$ /($10^{-3}\cdot$cm$^3\cdot$mol$^{-1}\cdot$K$^{-1}$) |
|---|---|---|---|---|---|
| | Experimental | PFP | | | |
| IB + $n$-heptane | -6.14 | -5.53 | 4.18 | -5.46 | -4.25 |
| IB + $n$-decane | -1.80 | -0.21 | 3.86 | -2.82 | -1.25 |
| IB + $n$-dodecane | -0.62 | 1.29 | 3.75 | -1.87 | -0.59 |
| IB + $n$-tetradecane | -0.48 | 2.31 | 3.87 | -1.27 | -0.29 |

**Figure S1**

Excess molar volume ($V_m^E$) of (iodobenzene (1) + *n*-alkane (2)) liquid mixtures as a function of the iodobenzene mole fraction ($x_1$) at temperature $T$ and pressure $p$ = 0.1 MPa. Symbols, experimental values: (□) $T$ = 288.15 K, (◇) $T$ = 293.15 K, (△) $T$ = 298.15 K, (×) $T$ = 303.15 K, (○) $T$ = 308.15 K. Solid lines are calculated with equations (10-11) using the coefficients from Table 7. (a) *n*-heptane, (b) *n*-decane.

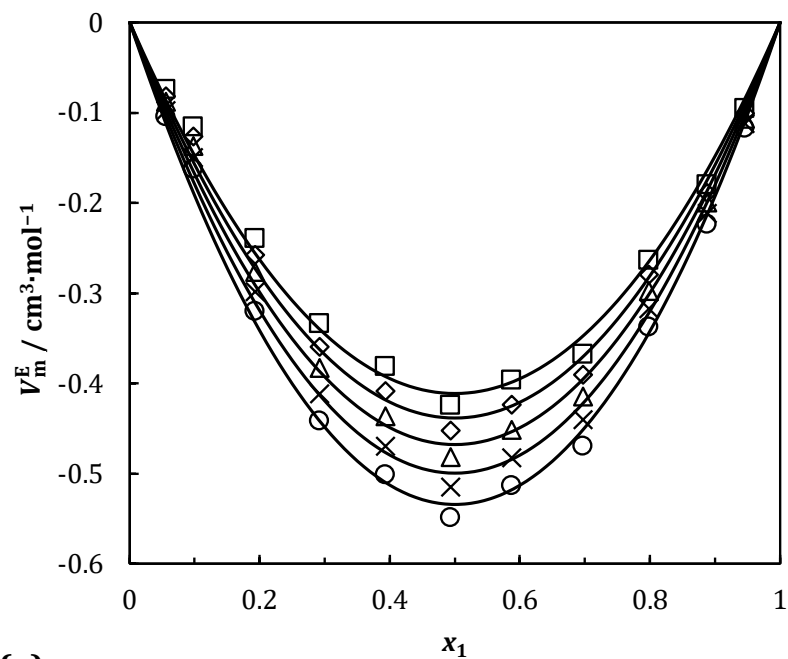

(a)

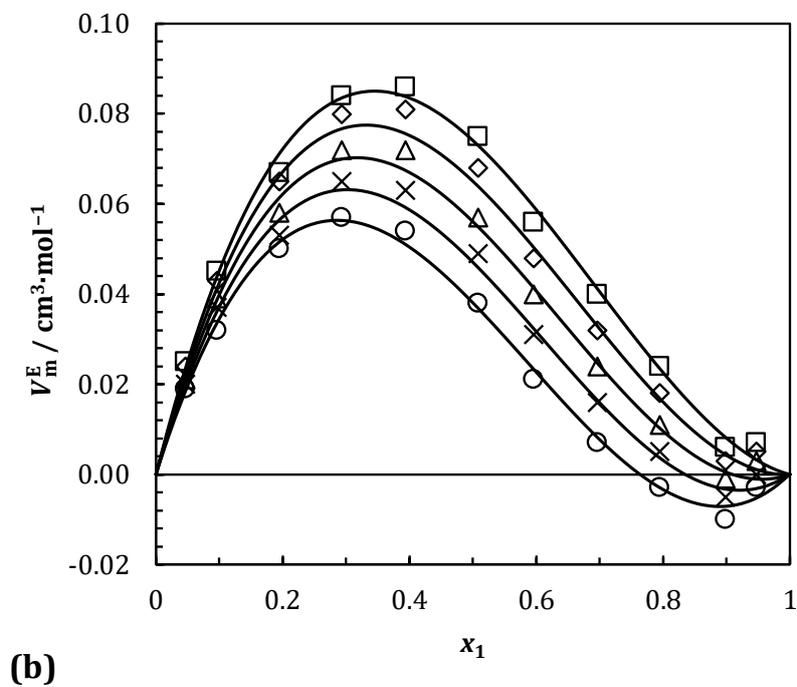

(b)

**Figure S2**

Temperature derivative of the excess molar volume $\left(\partial V_m^E/\partial T\right)_p$ of (iodobenzene (1) + *n*-heptane (2)) as a function of the iodobenzene mole fraction $(x_1)$ at temperature $T$ and pressure $p$ = 0.1 MPa: (solid line) $T$ = 288.15 K, (dotted line) $T$ = 293.15 K, (short-dashed line) $T$ = 298.15 K, (long-dashed line) $T$ = 303.15 K, (square dotted line) $T$ = 308.15 K. The calculations were performed from equations (10-11) using the coefficients from Table 7.

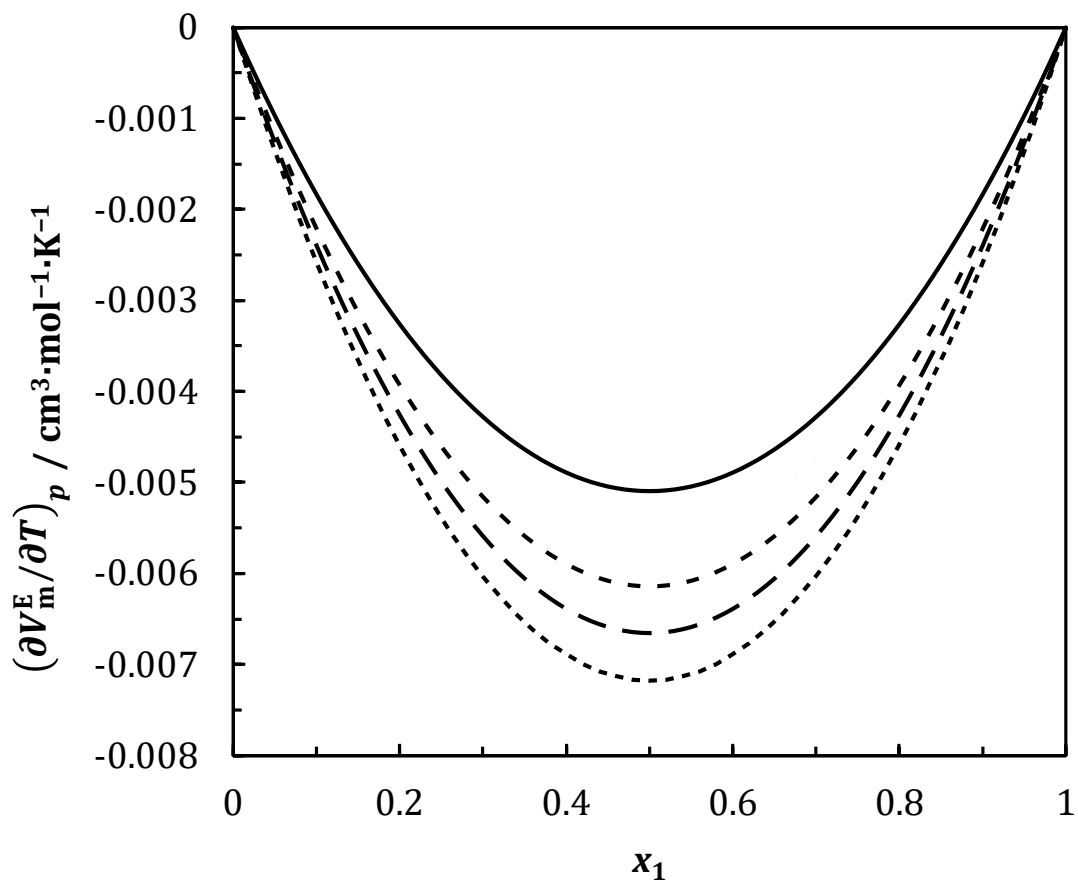

**Figure S3**

Contributions to the excess molar volume ($V_m^E$), obtained from the application of the Prigogine-Flory-Patterson (PFP) model to (iodobenzene (1) + n-alkane (2)) liquid mixtures at equimolar composition, temperature $T$ = 298.15 K and pressure $p$ = 0.1 MPa, as functions of the number of C atoms of the n-alkane ($n$). Symbols: (□) $V_{m,int}^E$, interactional term; (◇) $V_{m,p^*eff}^E$, $p^*$ effect term; (×) $V_{m,curv}^E$, curvature term; (○) $V_m^E$. Lines are drawn to aid the eye of the reader.

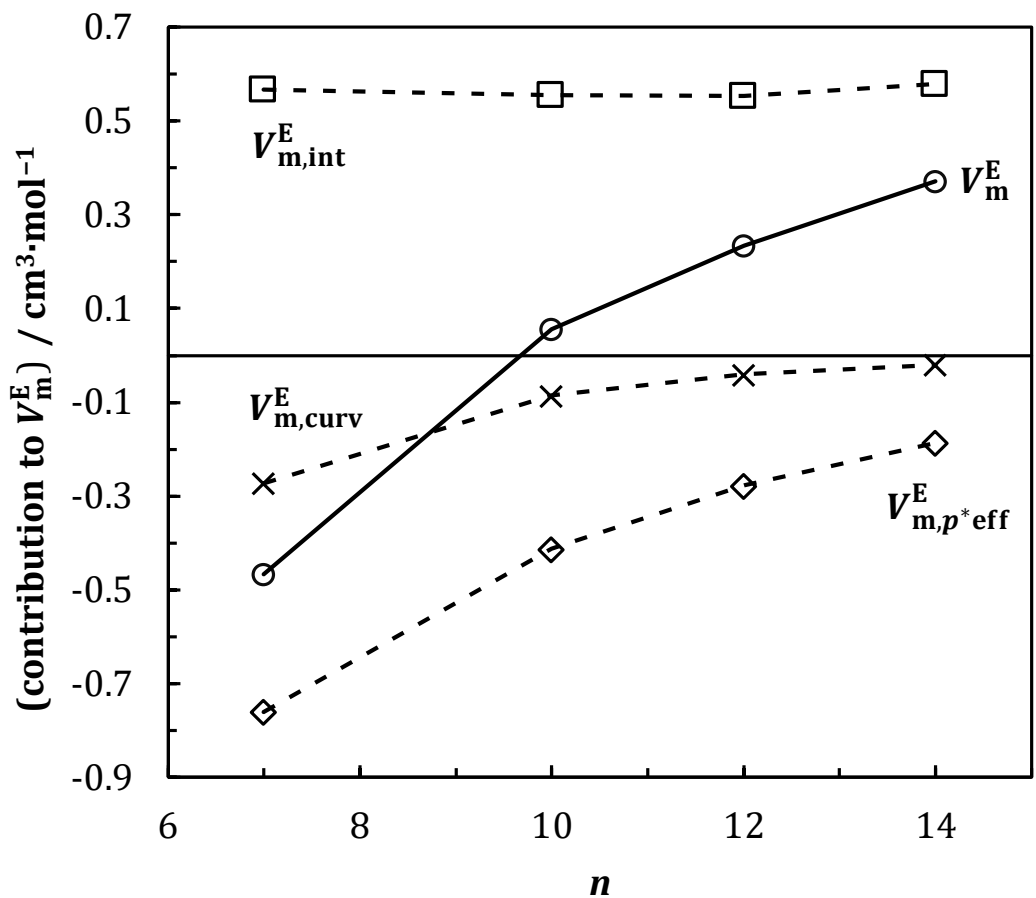

**References for supplementary material**